\documentclass[journal]{IEEEtran}
\usepackage{graphicx,amsmath,amssymb,amsfonts,subfigure}
\usepackage{verbatim} 

\begin{document}

\title{Random Matrix Model for Nakagami-Hoyt Fading}
        
\author{\IEEEauthorblockN{Santosh Kumar and Akhilesh Pandey}

\thanks{The work of S. Kumar was supported by CSIR India through fellowship.}
\thanks{The authors are with School of Physical Sciences, 
Jawaharlal Nehru University, New Delhi 110067, India 
(email: skumar.physics@gmail.com; ap0700@mail.jnu.ac.in).}}

\maketitle

\begin{abstract}
Random matrix model for the Nakagami-\textit{q} (Hoyt) fading in multiple-input multiple-output (MIMO) communication channels with arbitrary number of transmitting and receiving antennas is considered. The joint probability density for the eigenvalues of $\mathbf{H}^\dag \mathbf{H}$ (or $\mathbf{H}\mathbf{H}^\dag$), where $\mathbf{H}$ is the channel matrix, is shown to correspond to the Laguerre crossover ensemble of random matrices and is given in terms of a Pfaffian. Exact expression for the marginal density of eigenvalues is obtained as a series consisting of associated Laguerre polynomials. This is used to study the effect of fading on the Shannon channel capacity. Exact expressions for higher order density correlation functions are also given which can be used to study the distribution of channel capacity.
\end{abstract}

\begin{IEEEkeywords}
Channel Capacity, Fading distributions, Hoyt distribution, Laguerre crossover ensemble, MIMO channels, Nakagami-\textit{q} distribution, Random matrices. 
\end{IEEEkeywords}

\IEEEpeerreviewmaketitle

\section{Introduction}
\IEEEPARstart{T}{he} propagation of radio waves through wireless channels is a complex phenomenon. Various unavoidable effects associated with the multipath fading make it extremely difficult to formulate an exact mathematical description. However, there exist a number of simple and accurate statistical models for fading channels. These take into account the particular propagation environment and the underlying communication scenario \cite{SA1, SA2}. 

It has been shown that the use of multiple transmitting and receiving antennas leads to significant increase in the spectral efficiency of wireless communication systems \cite{FG1,Tltr}. This idea has been pursued vigorously in the last ten years \cite{FG1}-\cite{SL2}. There are now available a number of powerful results for the multiple-input multiple-output (MIMO) communication systems. The mathematical model for MIMO communication involves the channel matrix $\mathbf{H}$, whose entries characterize the paths between the antennas at the transmitting and receiving ends. Many quantities of interest, which measure performance of the MIMO system, follow from the eigenvalue statistics of Hermitian matrix $\mathbf{W}=\mathbf{H}^\dag \mathbf{H}$ (or $\mathbf{H}\mathbf{H}^\dag$). $\mathbf{H}$ is, in general, a rectangular random matrix with real or complex entries, its dimensions being $N_r\times N_t$. Here $N_r$ and $N_t$ are the numbers of receiving and transmitting antennas respectively. The choice of probability distribution for $\mathbf{H}$ depends on the propagation environment and communication scenario. When the matrix elements of $\mathbf{H}$ are independent and identically distributed (iid) Gaussian variables with zero mean, the ensemble of $\mathbf{W}$ is referred to as the Wishart or equivalently Laguerre ensemble \cite{Wish,Wilks,Edel,RM,PG,GP}. It has also been referred to as the Dyson ensemble \cite{Dys,RMP}.

The performance of MIMO systems with iid channels has been investigated extensively. The assumption of iid channels holds good when there is sufficient spacing within the antennas at either ends. Two important examples are the Rayleigh and Rician fadings with uncorrelated channels, where the matrix elements of $\mathbf{H}$ are taken to be iid complex Gaussians with zero or non-zero means respectively. These fading models apply to the cases with absence or presence of a strong direct line of sight in the propagation paths between the transmitting and receiving ends. Telatar \cite{Tltr} has considered the uncorrelated Rayleigh fading in great detail. His paper has been followed by many authors; see for example \cite{WG,VS,RP,SL1}. Rician fading with uncorrelated channels has been studied in \cite{CD,JP}. In recent investigations the condition of iid channels has been relaxed and the effects of spatial correlations has been considered \cite{KA2,Mlr, SRS, CWZ}. This is needed to take account of closely spaced antennas at either or both the transmitting and receiving ends. 

The general problem of multichannel communication has been considered by Simon, Moustakas and others \cite{SM,SMM,GTN,MA} by group theoretical methods to obtain exact results for both uncorrelated and correlated cases. See also \cite{KA1} and \cite{OPF} respectively for other ensembles and another approach to the problem of fading in MIMO channels.

Most of the research related to MIMO capacity has focused mainly on Rayleigh and Rician distributions. However, there are other classes of fading distributions which serve as better models under certain circumstances. An example is the Nakagami-$m$ distribution which contains Rayleigh, one-sided Gaussian and uniform distribution on unit circle as special cases \cite{Nak,MS,FLC2}. Another example is the Weibull distribution which has been considered in \cite{SK}.

Our purpose in this paper is to study Nakagami-$q$ distribution \cite{SA1,SA2,Nak,Hoyt,FLC1} with iid channels.  This model was proposed by Nakagami as an approximation to the above mentioned Nakagami-$m$ distribution for certain range of $m$ values \cite{SA1,SA2,Nak,FLC2}. It has been investigated independently by Hoyt \cite{Hoyt}, and hence is also referred to as Nakagami-Hoyt fading. Under this fading the elements of $\mathbf{H}$ are taken to be complex iid Gaussians with zero mean, but in general having different variances for real and imaginary parts. It spans the range from one-sided Gaussian to Rayleigh distribution as the value of parameter $q$ is varied from 0 to 1. Note here that one-sided Gaussian fading occurs in the limit when variance of imaginary (or real) part is zero, whereas Rayleigh case is obtained when the variances of real and imaginary parts are equal. This fading is observed on satellite links which are subject to strong ionospheric scintillation \cite{SA1,SA2,Chy}. It has also been used to analyze the cellular mobile radio systems \cite{ATB,YWP}. 

In the context of random matrix theory Rayleigh fading corresponds to the Laguerre unitary ensemble (LUE) \cite{Edel,RM,PG,GP} of random matrices, whereas one sided Gaussian fading corresponds to the Laguerre orthogonal ensemble (LOE) of random matrices \cite{Edel,RM,PG,GP}. Thus Nakagami-$q$ fading, in general, corresponds to a random matrix ensemble which interpolates between LOE and LUE. This crossover ensemble, along with the more general Jacobi crossover ensembles, was recently studied by the present authors \cite{SKP1,SKP2}. In this paper we apply the LOE-LUE crossover results \cite{SKP1} to solve the problem of Nakagami-$q$ fading. The crossover results follow from the Brownian motion formalism of random matrices \cite{Dys,Pandey}. This approach is equivalent to the group theoretical formalism mentioned above and has been briefly described in Appendix A.

The Nakagami-$q$ fading results for $N_t=N_r=2$ were obtained recently by Fraidenraich \textit{et al.} \cite{FLC1}. These results were presented in terms of complicated multiple integrals. In the present work we obtain exact results in compact forms for all $N_t, N_r$. We obtain the JPD of eigenvalues in terms of a Pfaffian \cite{RM} (See Appendix B) from which all density correlation functions can be derived. In particular the level density, which is proportional to the marginal density, is given in terms of a series consisting of associated Laguerre polynomials. We use this to study the mean or ergodic capacity of MIMO channels for Nakagami-$q$ fading. The higher order correlation functions can be used to study distribution of the channel capacity \cite{SS,GWB,KHY,YMLJ}. These exact results serve as interesting addition to already existing rich literature in this field.

The paper is organized as follows. In Sec. \ref{secNq} we describe the Nakagami-$q$ distribution. Sec. \ref{secMIMO} deals with the Nakagami-$q$ fading in MIMO systems and gives the LOE-LUE crossover results. Some of the details of the derivation are given in Appendices A, B, C and D. In Sec. \ref{secSCC} the effect of this fading on Shannon channel capacity is investigated. We conclude with some general remarks in Sec. \ref{secConc}.

\section{Nakagami-$q$ Distribution }
\label{secNq}

The instantaneous signal in Nakagami-$q$ fading is modelled as the complex Gaussian quantity
\begin{equation}
\label{Z}
Z=X+\mbox{j}\,Y=Re^{\mbox{j}\theta},
\end{equation}
where $X, Y$ are zero mean Gaussians with variances $\sigma_X^2$ and $\sigma_Y^2$. The distribution of the signal envelope $R=|Z|$ is obtained as \cite{Nak,Hoyt}
\begin{equation}
\label{Nq}
p_R(R)=\frac{(1+q^2)R}{q\Omega}\,e^{-\frac{(1+q^2)^2}{4q^2\Omega}R^2}I_0\bigg(\frac{(1-q^4)}{4q^2\Omega}R^2\bigg),
\end{equation}
where $I_0(\cdot)$ represents the zeroth order modified Bessel function of the first kind. The distribution of phase $\theta$ is given by
\begin{equation}
p_\theta(\theta)=\frac{\sigma_X \sigma_Y}{2\pi(\sigma_X^2\sin^2\theta+\sigma_Y^2\cos^2\theta)},~~~-\pi\leq\theta<\pi. 
\end{equation}
In (\ref{Nq}) $q\in[0,1]$ is the Nakagami-Hoyt fading parameter, and $\Omega=\mathbb{E}[R^2]=\sigma_X^2+\sigma_Y^2$. $\mathbb{E}[\cdot]$ is the expectation operator and represents the statistical average. The parameter $q$ is given in terms of the variances $\sigma_X^2$ and $\sigma_Y^2$ as
\begin{equation}
\label{NHq}
q=\begin{cases}
\sigma_Y/\sigma_X, & \sigma_X \geq \sigma_Y\\
  \sigma_X/\sigma_Y, & \sigma_Y \geq \sigma_X ~~~.
\end{cases} 
\end{equation}
Thus we have $q=0$ for $\sigma_Y=0$ (or $\sigma_X=0$), the other being nonzero, and $q=1$ for $\sigma_X=\sigma_Y$. These correspond respectively to one sided Gaussian and Rayleigh fadings. Note that the $q=0$ case is referred to as one-sided Gaussian because the signal amplitude (\ref{Nq}) corresponds to that of $|X|$ or $|Y|$.

For definiteness we choose $\sigma_X \geq \sigma_Y$. Note then that the two cases in (\ref{NHq}) correspond to the distribution of $Z$ and $Z/\mbox{j}$\,. The Brownian motion parameter $\tau$ \cite{SKP1} which governs the LOE-LUE crossover (see Appendix B) is related to the quantities defined here in the following way:
\begin{equation}
\label{tau}
e^{-\tau}=\frac{\sigma_X^2-\sigma_Y^2}{\sigma_X^2+\sigma_Y^2}=\frac{1-q^2}{1+q^2},
\end{equation}
which is also the parameter $b$ in \cite{FLC1}. Alternatively we can write
\begin{equation}
\label{vxvy}
\sigma_X^2=\left(\frac{1+e^{-\tau}}{2}\right)\Omega,~~~\sigma_Y^2=\left(\frac{1-e^{-\tau}}{2}\right)\Omega.
\end{equation}
The crossover results in \cite{SKP1} correspond to the choice $\Omega=1/2$. Also, $\tau=0$ corresponds to LOE whereas $\tau\rightarrow\infty$ corresponds to LUE. These limits lead to one-sided Gaussian ($q=0$) and Rayleigh ($q=1$) fadings respectively. 

\section{MIMO Channel Model and LOE-LUE Crossover}
\label{secMIMO}

We consider a single user Gaussian channel with $N_t$ antennas at the transmitter end and $N_r$ antennas at the receiver end. The MIMO channel can then be modelled as \cite{Tltr}
\begin{equation}
\label{MIMO}
\mathbf{y}=\mathbf{H}\,\mathbf{x}+\mathbf{n}~.
\end{equation}
Here $\mathbf{x}$ and $\mathbf{y}$ represent the ($N_t$ and $N_r$ dimensional) transmitted and received signal vectors respectively. $\mathbf{n}$ is complex Gaussian noise vector with zero mean ($\mathbb{E}[\mathbf{n}]=0)$ and unit variance ($\mathbb{E}[\mathbf{n}\mathbf{n}^\dag]=\mathbf{I}_{N_r}$). $\mathbf{H}$ is $N_r\times N_t$ dimensional channel matrix with iid entries $H_{jk}$ distributed as the random variable $Z$ given in (\ref{Z}). Equivalently $\mathbf{H}$ can also be written as 
\begin{equation}
\label{Hmat}
\mathbf{H}=\mathbf{H}_X+\mbox{j}\,\mathbf{H}_Y,
\end{equation}
where $\mathbf{H}_X$ and $\mathbf{H}_Y$ are independent $N_r\times N_t$ real matrices with zero mean iid Gaussian entries with variances $\sigma_X^2$ and $\sigma_Y^2$ respectively. Now consider $N$=min$(N_t,N_r)$ and $M$=max$(N_t,N_r)$. The Hermitian matrix
\begin{equation}
\label{wishart}
\mathbf{W}=\begin{cases}
\mathbf{H}^\dag \mathbf{H}\mbox{ for } N_r\ge N_t\\
\mathbf{H}\mathbf{H}^\dag \mbox { for } N_r < N_t
\end{cases}
\end{equation}
is therefore $N\times N$ dimensional. 

As mentioned in the introduction the matrix model given by (\ref{Hmat}) maps to the random matrix problem of LOE-LUE crossover \cite{SKP1}. The JPD of eigenvalues $\{\lambda\}\equiv (\lambda_1,...,\lambda_N$) of $\mathbf{W}$ for this crossover can be obtained using the Brownian motion model of random matrices \cite{Dys,Pandey}. Proofs are given in Appendices A and B \footnote{Appendix A also gives the JPD of eigenvalues for uncorrelated Rician MIMO channels \cite{SMM} which arises as an intermediate step in our derivation.}. We find that \cite{SKP1}
\begin{equation}
\label{JPD}
P(\{\lambda\};\Omega;\tau)=\frac{2^m e^{N(N-1)\tau/2}}{(2\Omega)^{N(N+1)/2}}C_{N}^{(0)}\Delta_N\mbox{Pf}[F_{j,k}^{(\tau)}]\prod_{l=1}^N w_a\Big(\frac{\lambda_l}{2\Omega}\Big).
\end{equation}
Here Pf($\mathbf{B}$) represents the Pfaffian of an even-dimensional antisymmetric matrix $\mathbf{B}$ \cite{RM} (see Appendix B for definition),
\begin{equation}
\label{CNzero}
C_N^{(0)}=\frac{\pi^{N/2}}{2^N}
\prod_{k=1}^N\frac{1}{\Gamma(\frac{k}{2}+1)\Gamma(\frac{k}{2}+\frac{2a+1}{2})} 
\end{equation}
is the normalization,
\begin{equation}
\Delta_N\equiv\Delta_N(\lambda_1,...,\lambda_N)=\prod_{1\le j<k\le N}(\lambda_j-\lambda_k) 
\end{equation}
is the Vandermonde determinant and 
\begin{equation}
w_a(x)=x^a e^{-x},~~~~~~~~~0\leq x<\infty,  
\end{equation}
is the associated Laguerre weight function. The parameter $a$ is defined by 
\begin{equation}
2a+1=|N_t-N_r|.
\end{equation}
$\mathbf{F}^{(\tau)}$ is a $2m$-dimensional antisymmetric matrix with $2m=N$ or $N+1$ depending on whether $N$ is even or odd. The matrix elements $F_{j,k}^{(\tau)}$ are obtained from their $\tau=0$ counterparts using the one-body operators introduced in \cite{SKP1} (See Appendices B and C). For $j,k=1,...,N$ we have
\begin{equation}
\label{Fjk_e}
F_{j,k}^{(\tau)}=\mathcal{G}^{(\tau)}\Big(\frac{\lambda_j}{2\Omega},\frac{\lambda_k}{2\Omega}\Big).
\end{equation}
For odd $N$, we have in addition
\begin{equation}
\label{Fjk_o}
F_{j,N+1}^{(\tau)}=-F_{N+1,j}^{(\tau)}=\omega^{(\tau)}\Big(\frac{\lambda_j}{2\Omega}\Big)(1-\delta_{j,N+1}).
\end{equation}
With $L_\mu^{(\alpha)}(x)$ representing associated Laguerre polynomials, $\mathcal{G}^{(\tau)}(x,y)$ and $\omega^{(\tau)}(x)$ are given by
\begin{eqnarray}
\label{Gxyt}
\nonumber
&&\!\!\!\mathcal{G}^{(\tau)}(x,y)\\
\nonumber
&=&2\, w_{a+1}(x)w_{a+1}(y)\sum_{\mu=0}^{\infty}\sum_{\nu=\mu}^{\infty}e^{-(2\mu+2\nu+1)\tau}\kappa_{\mu,\nu}^{(a)}\\
\nonumber
&\times&\!\!\!\!\!\Big[L_{2\mu}^{(2a+1)}(2x)L_{2\nu+1}^{(2a+1)}(2y)-L_{2\nu+1}^{(2a+1)}(2x)L_{2\mu}^{(2a+1)}(2y)\Big]\\
\nonumber
\!\!&=&\!\!2\,w_{a+1}(x)w_{a+1}(y)\sum_{\mu=0}^{\infty}\sum_{\nu=0}^{\mu}e^{-(2\nu+2\mu+1)\tau}\kappa_{\nu,\mu}^{(a)}\\
\nonumber
&\times&\!\!\!\!\!\Big[L_{2\nu}^{(2a+1)}(2x)L_{2\mu+1}^{(2a+1)}(2y)-L_{2\mu+1}^{(2a+1)}(2x)L_{2\nu}^{(2a+1)}(2y)\Big],\\ 
\end{eqnarray}
where
\begin{equation}
\kappa_{\mu,\nu}^{(a)}=\frac{\Gamma(\mu+\frac{1}{2})\Gamma(\nu+1)}{\Gamma(\mu+a+\frac{3}{2})\Gamma(\nu+a+2)},
\end{equation}
and
\begin{equation}
\label{OMGxt}
\omega^{(\tau)}(x)=\,w_{a+1}(x)\sum_{\mu=0}^{\infty}\frac{e^{-2\mu\tau}\Gamma(\mu+\frac{1}{2})}{\Gamma(\mu+a+\frac{3}{2})}L_{2\mu}^{(2a+1)}(2x).
\end{equation}

\begin{figure*}[!t]
\centerline{\subfigure[$N_t=2, N_r=2$]{\includegraphics[width=3.4in]{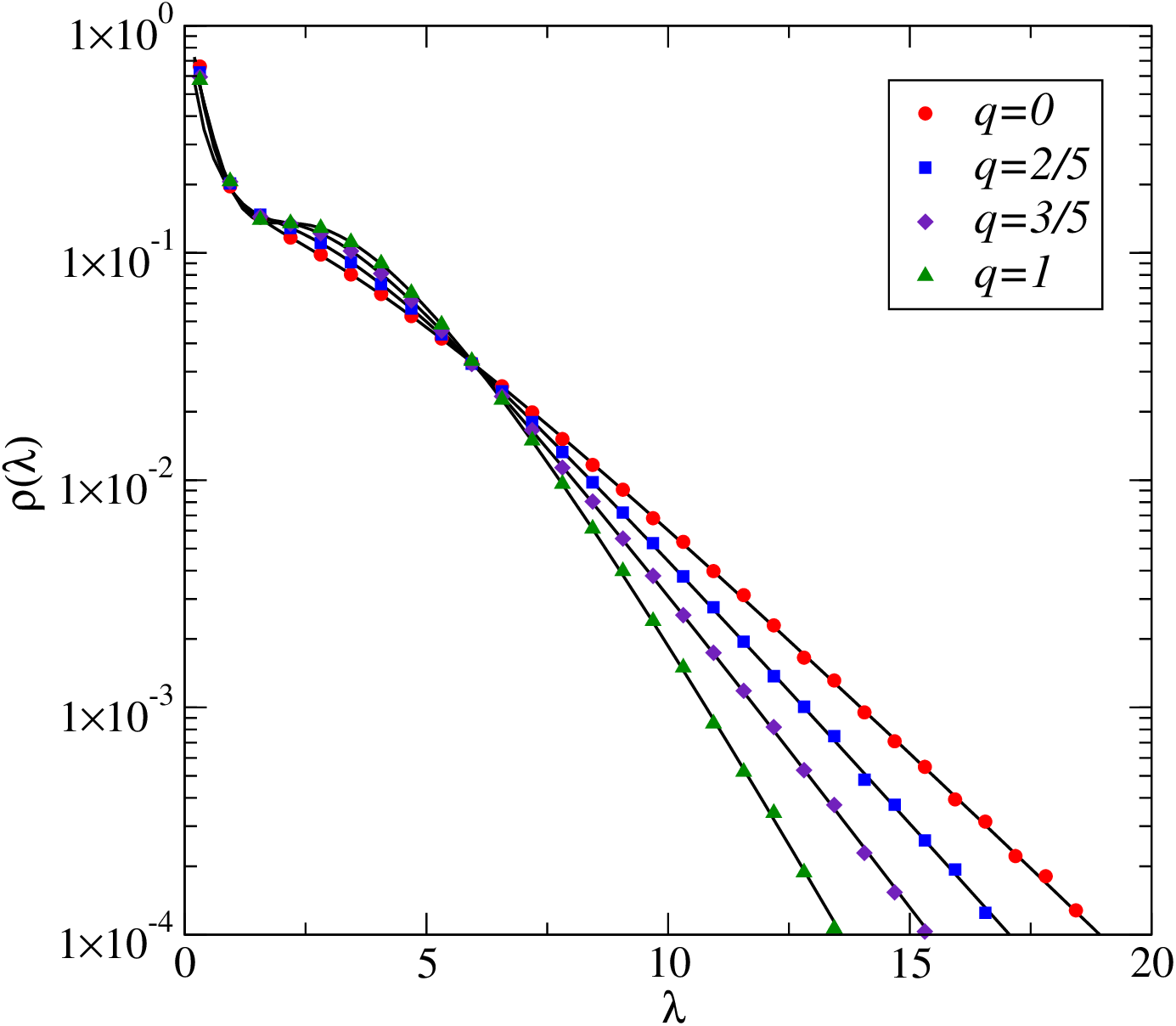}
\label{fig_first_case}}
\hfil
\subfigure[$N_t=3, N_r=6$]{\includegraphics[width=3.4in]{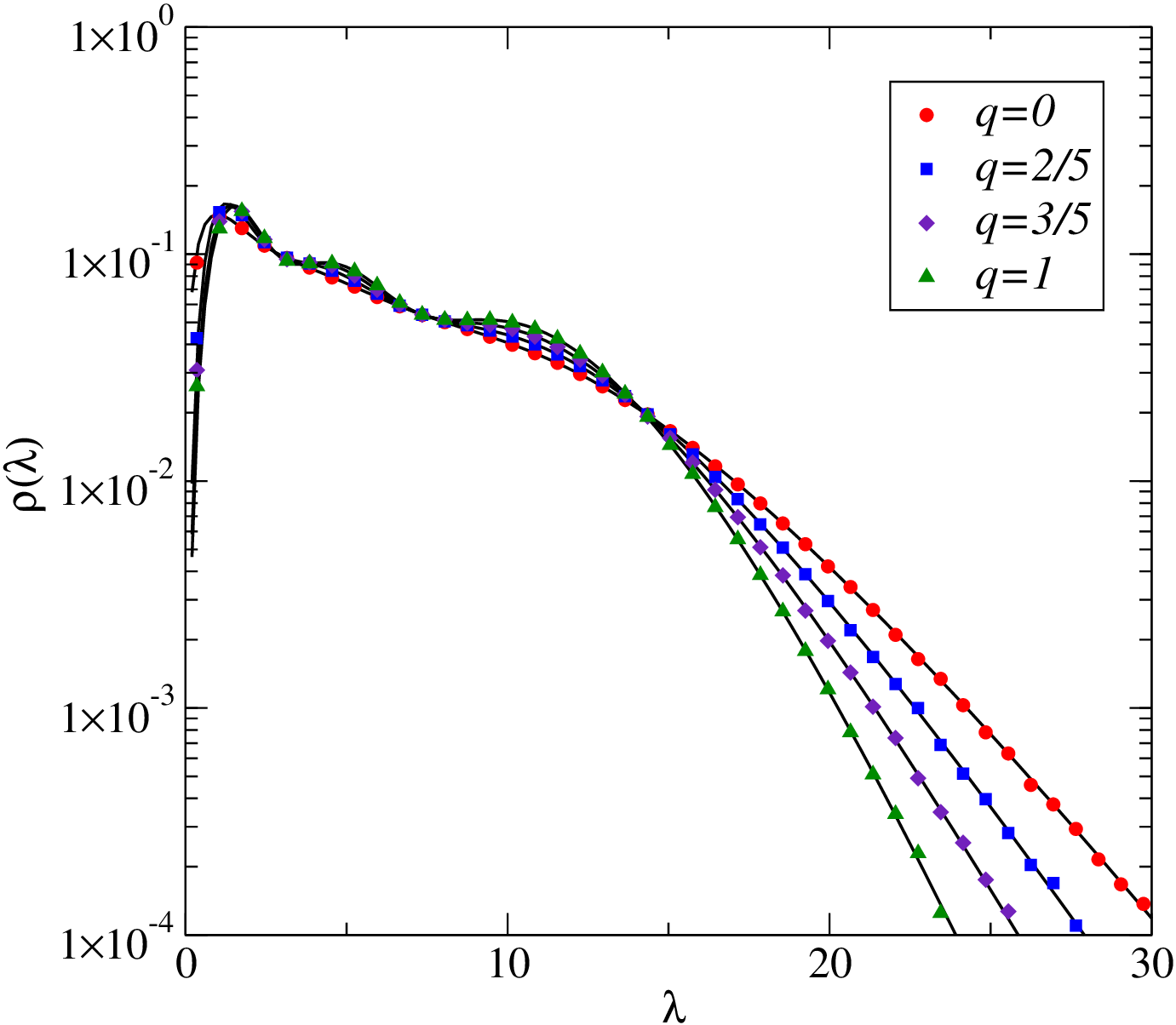}
\label{fig_second_case}}
}
\label{fig1a}
\end{figure*}
\begin{figure*}[!ht]
\centerline{\subfigure[$N_t=4, N_r=15$]{\includegraphics[width=3.4in]{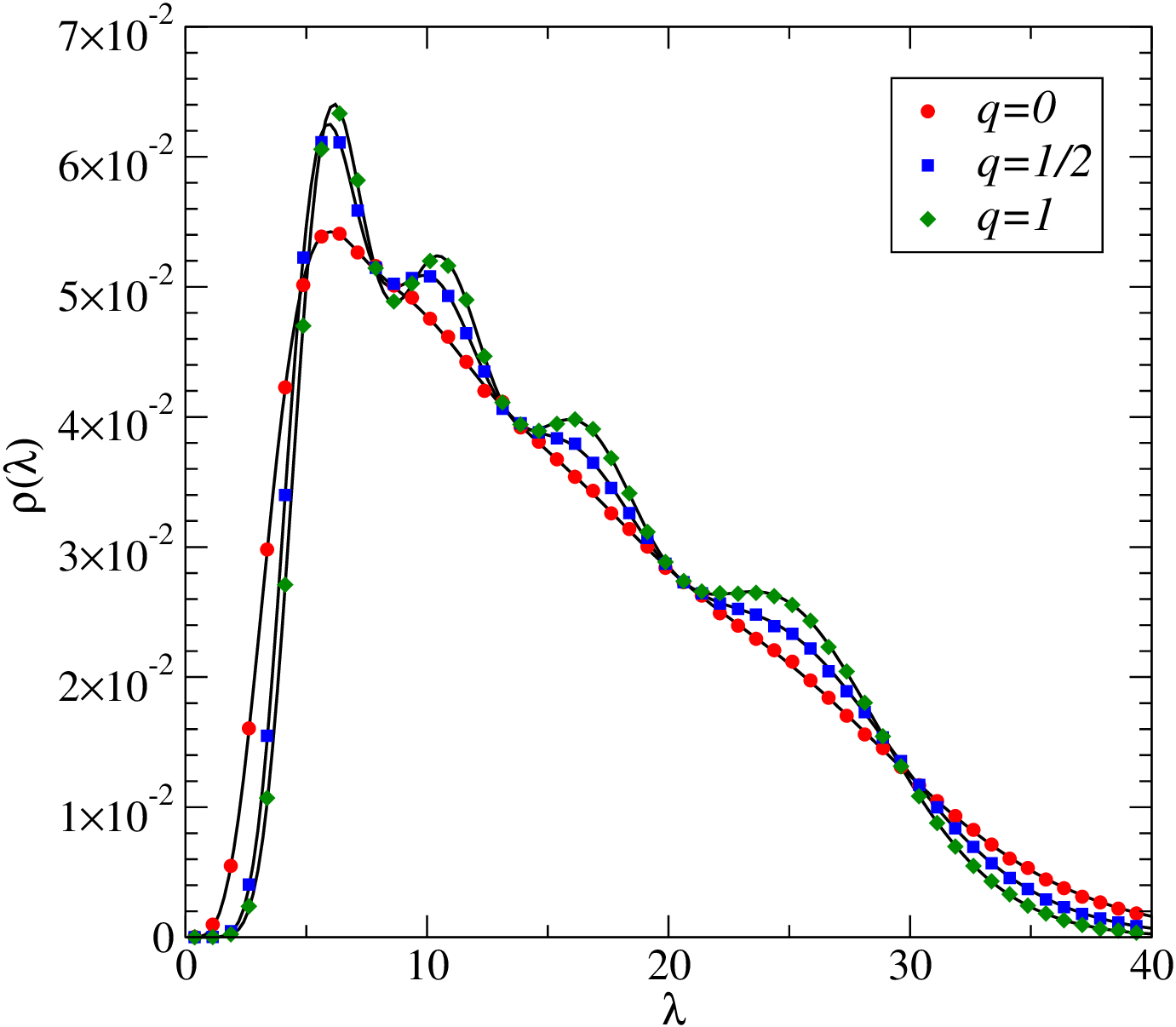}
\label{fig_third_case}}
\hfil
\subfigure[$N_t=16, N_r=16$]{\includegraphics[width=3.4in]{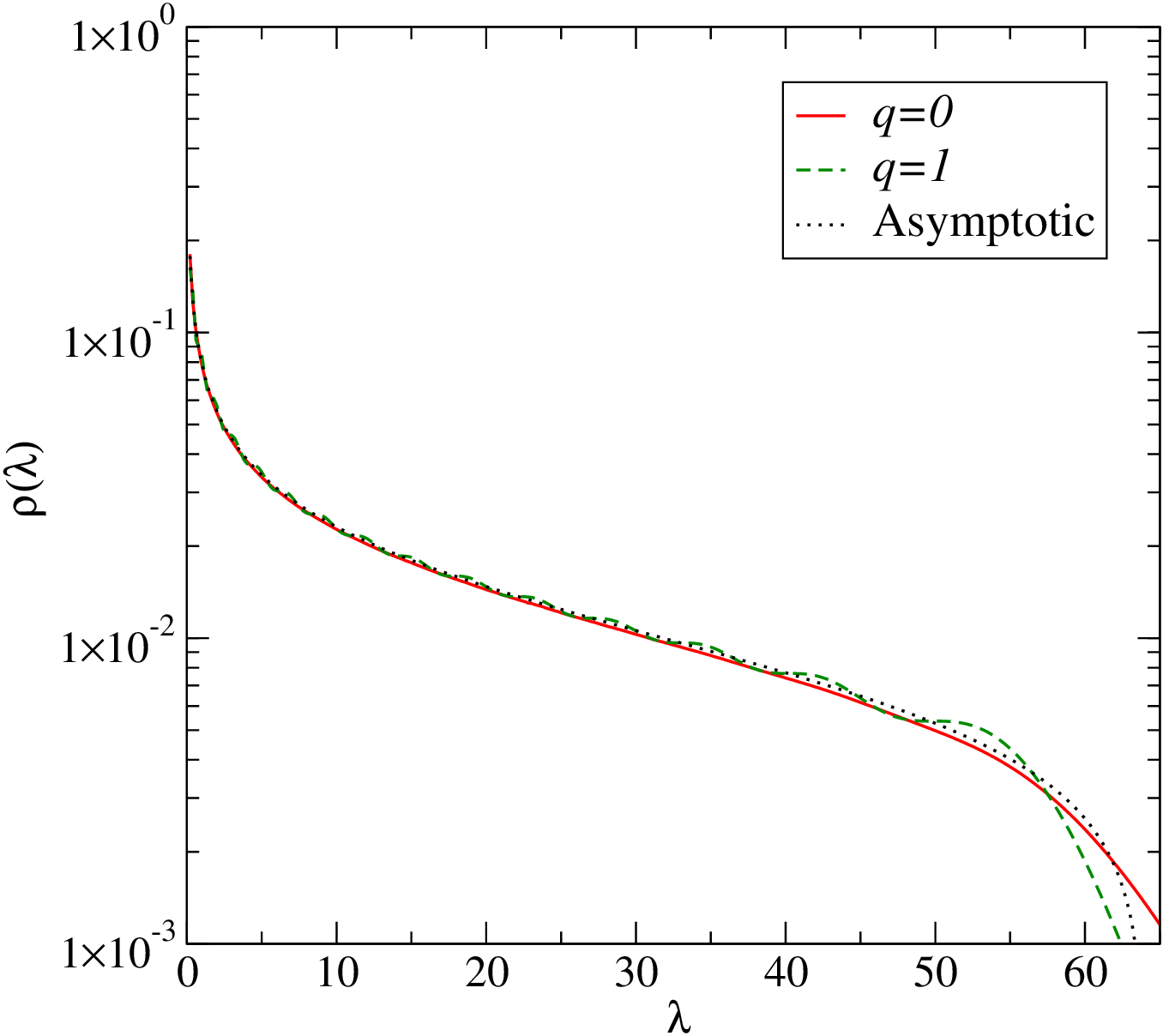}
\label{fig_fourth_case}}
}
\caption{Marginal density of eigenvalues $\rho(\lambda;\Omega;\tau)$ for $\Omega=1$ and several combinations of $N_t, N_r$. The inset gives the $q$ values which is related to $\tau$ by (\ref{tau}). The lines are theoretical predictions whereas the symbols are simulation results. In (d) we have avoided simulation results for clarity and instead given the asymptotic result calculated using Eq. (\ref{Asym}).}
\label{fig1b}
\end{figure*}

\noindent
The two representations of $\mathcal{G}^{(\tau)}(x,y)$ appearing in (\ref{Gxyt}) follow from each other by changing the order of the two summations; see appendix D. Also note that $\mathcal{G}^{(\tau)}(x,y)$ is an antisymmetric function of $x$ and $y$. As explained in \cite{SKP1,SKP2}, these results valid for arbitrary $\tau$ derive from the $\tau=0$ results for which
\begin{equation}
\label{Gxy0}
\mathcal{G}^{(0)}(x,y)=\frac{1}{2}\mbox{sgn}(x-y),
\end{equation}
and
\begin{equation}
\label{OMGx0}
\omega^{(0)}(x)=\frac{1}{2}.
\end{equation}
For $\tau=0~(q=0)$ and $\tau=\infty~(q=1)$ equation (\ref{JPD}) reduces to the following well known results for the LOE and LUE respectively \cite{Wish,Wilks,Edel,RM,PG,GP,Dys},
\begin{equation}
\label{JPDzero}
P(\{\lambda\};\Omega;0)=\frac{1}{(2\Omega)^{N(N+1)/2}}C_N^{(0)} |\Delta_N|\prod_{l=1}^N w_a\Big(\frac{\lambda_l}{2\Omega}\Big), 
\end{equation}
\begin{equation}
\label{JPDinf}
P(\{\lambda\};\Omega;\infty)=\frac{1}{\Omega^{N^2}}C_N^{(\infty)} (\Delta_N)^2\prod_{l=1}^N w_{2a+1}\Big(\frac{\lambda_l}{\Omega}\Big).
\end{equation}
The normalization $C_N^{(0)}$ is given by Eq. (\ref{CNzero}) and $C_N^{(\infty)}$ by
\begin{equation}
\label{CNinf}
C_N^{(\infty)}=\prod_{k=1}^N \frac{1}{\Gamma(k+1)\Gamma(k+2a+1)}.
\end{equation}
Equation (\ref{JPDzero}) applies to the one-sided Gaussian fading, whereas (\ref{JPDinf}) applies to the Rayleigh fading. The latter is considered in \cite{Tltr} with the choice $\Omega=1$.

The $n$-level correlation function defined by
\begin{eqnarray}
\label{Rn}
\nonumber
R_n(\lambda_1,...,\lambda_n;\Omega;\tau)\!\!\!&=&\!\!\!\frac{N!}{(N-n)!}\int_{0}^{\infty}\cdots\int_{0}^{\infty}\!\!P(\{\lambda\};\Omega;\tau)\\
&&~~~~~~~~~~~~~\times  d\lambda_{n+1}...d\lambda_N
\end{eqnarray}
can be expressed as a quaternion determinant involving certain two-point kernels, which in turn derive from skew-orthogonal polynomials and their dual functions \cite{PG, GP, SKP1, SKP2}. The explicit forms have been given in Appendix C. Since we are interested here in the calculation of channel capacity, which is a linear statistic on eigenvalues, we need $R_1(\lambda;\Omega;\tau)$ only. 
We find,
\begin{eqnarray}
\label{dens}
\nonumber
&&\!\!\!R_1(\lambda;\Omega;\tau)=R_1(\lambda;\Omega;\infty)~~~~\\
\nonumber
&+&\!\!\!\frac{1}{\Omega}w_{2a+1}\Big(\frac{\lambda}{\Omega}\Big)L_{N-1}^{(2a+1)}\Big(\frac{\lambda}{\Omega}\Big)\sum_{\nu=\frac{(N+c)}{2}}^\infty \!\!\!\frac{e^{-(2\nu+2-N-c)\tau}}{2^{2a+1}}\\
&\times&\!\!\! \frac{\,\Gamma\Big(\frac{N+1}{2}\Big)\,\Gamma(\nu+1-\frac{c}{2})}{\,\Gamma\Big(\frac{N+2a+1}{2}\Big)\,\Gamma(\nu+a+2-\frac{c}{2})}L_{2\nu+1-c}^{(2a+1)}\Big(\frac{\lambda}{\Omega}\Big),
\end{eqnarray}
where
\begin{equation}
c=N(\mbox{mod }2).
\end{equation}
A proof of (\ref{dens}) is outlined in Appendix D. In the limit $\tau\rightarrow\infty$ the second term in (\ref{dens}) vanishes. The first term gives the density for the Rayleigh case (LUE result), viz.
\begin{eqnarray}
\label{Dinf}
\nonumber
&&\!\!\!\!\!\!R_1(\lambda;\Omega;\infty)~~~~~~~~~~~~~~~~~~~~~~~~~~~~~~~~~~~~~~~~~~~~~~~~~~~\\
&=&\!\!\!\frac{1}{\Omega}w_{2a+1}\Big(\frac{\lambda}{\Omega}\Big)\!\!\sum_{\mu=0}^{N-1}\frac{\Gamma(\mu+1)}{\Gamma(\mu+2a+2)}\Big[L_\mu^{(2a+1)}\Big(\frac{\lambda}{\Omega}\Big)\Big]^2.
\end{eqnarray}
For $\tau=0$ (\ref{dens}) can be reduced to a finite sum involving incomplete Gamma function $\Gamma(s,x)=\int_x^\infty y^{s-1}e^{-y} dy$ along with the associated Laguerre polynomials, i.e.
\begin{eqnarray}
\label{Dzero}
\nonumber
&&\!\!\!\!\!R_1(\lambda;\Omega;0)=R_1(\lambda;\Omega;\infty)~~~~\\
\nonumber
&+&\!\!\!\!\frac{1}{2\Omega}w_a\Big(\frac{\lambda}{2\Omega}\Big)L_{N-1}^{(2a+1)}\Big(\frac{\lambda}{\Omega}\Big)\sum_{\mu=0}^N \chi_{N,\mu}^{(a)}\Gamma\Big(\mu+a+1,\frac{\lambda}{2\Omega}\Big)\\
&+&\!\!\!\frac{1}{2\Omega}w_a\Big(\frac{\lambda}{2\Omega}\Big)L_{N-1}^{(2a+1)}\Big(\frac{\lambda}{\Omega}\Big)\Big[c\,\xi_{N}^{(a)}-(1-c)\eta_{N}^{(a)}\Big]
\end{eqnarray}
with
\begin{equation}
 \chi_{N,\mu}^{(a)}=\frac{(-1)^\mu 2^{\mu+2a+1}(N+2a+1)\Gamma(N+1)}{\Gamma(\mu+1)\Gamma(\mu+2a+2)\Gamma(N-\mu+1)},
\end{equation}
\begin{equation}
\xi_{N}^{(a)}=\frac{\Gamma(\frac{N+1}{2})}{\Gamma(\frac{N+2a+1}{2})}, 
\end{equation}
and
\begin{equation}
\eta_{N}^{(a)}=\frac{2^{2a}\Gamma(N+1)\Gamma(\frac{N+2a+2}{2})}{\Gamma(N+2a+1)\Gamma(\frac{N+2}{2})}.
\end{equation}
Note that for $\tau\neq 0$ the series in (\ref{dens}) converges very rapidly because of the presence of exponential factor. For $\tau=0$ equation (\ref{Dzero}) is more useful for numerical considerations than (\ref{dens}).

For large $N$ the first term in (\ref{dens}) dominates over the second and so the density becomes independent of $\tau$ (or $q$) and is described by the Mar\u{c}enko-Pastur-Dyson density \cite{Dys,MP,Bai},
\begin{equation}
\label{Asym}
R_1(\lambda;\Omega)=\begin{cases}
           \dfrac{\sqrt{(\lambda_{\text{max}}-\lambda)(\lambda-\lambda_{\text{min}})}}{2\pi\Omega\lambda}, & \lambda_{\text{min}} \leq \lambda \leq \lambda_{\text{max}},\\
               ~~~~~~~0~, & \mbox{otherwise}.
              \end{cases} 
\end{equation}
Here $\lambda_{\text{min}}=M\Omega(1-\sqrt{N/M})^2$ and $\lambda_{\text{max}}=M\Omega(1+\sqrt{N/M})^2$.

The marginal density $\rho(\lambda;\Omega;\tau)=R_1(\lambda;\Omega;\tau)/N$ (normalized to 1) has been shown in Fig. 1 for different $N_t, N_r$ at various $q$ values. The theoretical predictions based on (\ref{dens}) and the results of numerical simulations for the matrix model given by (\ref{Hmat}) are in excellent agreement. In Fig. 1(d) for $N_t=N_r=16$ the density curves for $q=0$ and $q=1$ almost overlap and are well described by (\ref{Asym}).

We would like to remark here that it is difficult to analytically demonstrate the equivalence of $N_t=N_r=2$ results in \cite{FLC1} with our corresponding results. However the agreement of theoretical predictions with numerical simulations in both cases show that our results are equivalent. 

\begin{figure}[!t]
\centering
\includegraphics[width=3in]{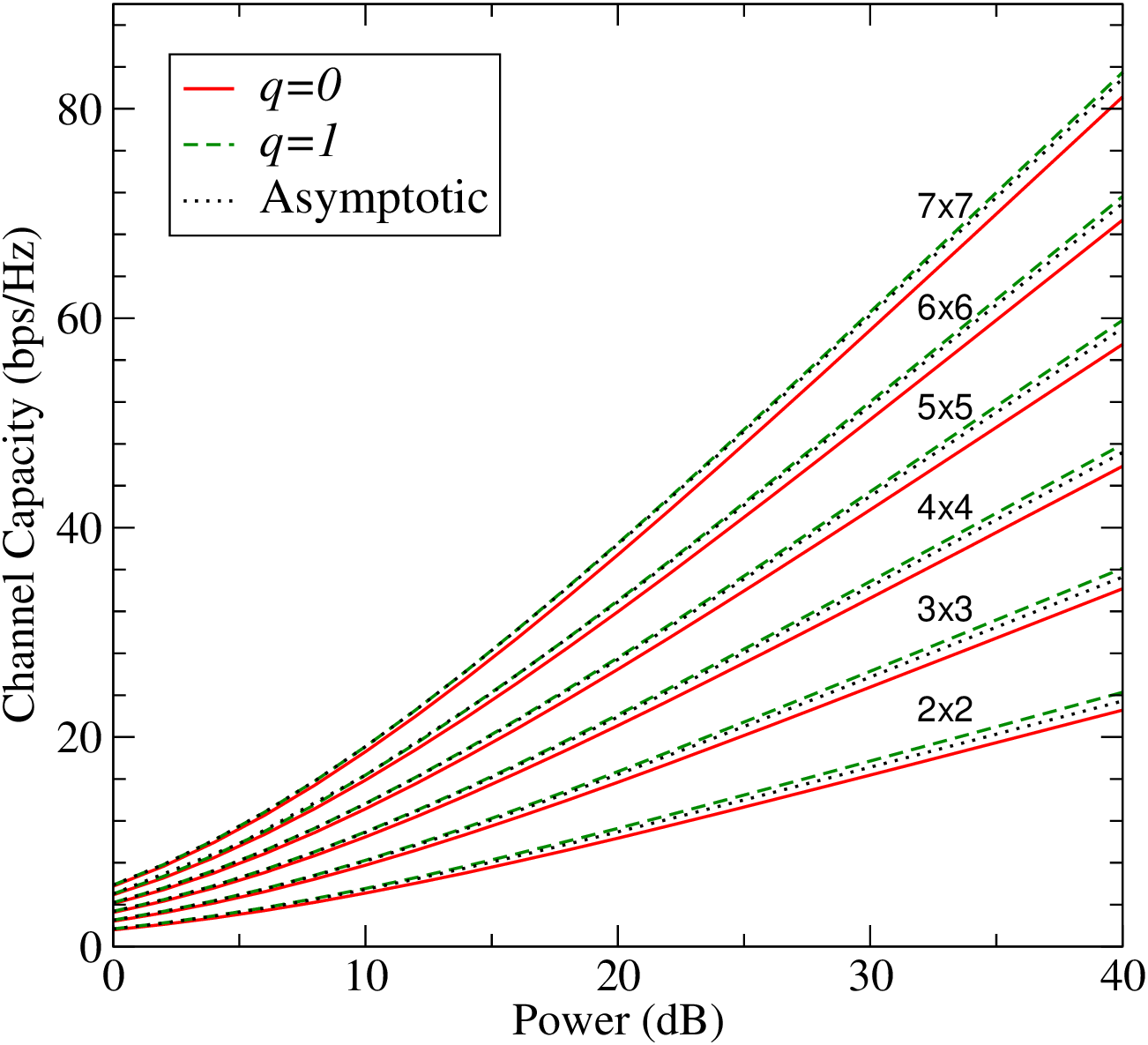}
\caption{Channel Capacity vs Power for $q=0$ and $q=1$ for several $N_t\times N_r$ values with $\Omega=1$.}
\label{ChCap}
\end{figure}

\section{Shannon Channel Capacity}
\label{secSCC}

In the absence of channel knowledge at the transmitter end the ergodic MIMO channel capacity is given as \cite{Tltr}
\begin{eqnarray}
\label{ccap}
\nonumber
\mathcal{C}\!\!\! &=&\!\!\! \mathbb{E}\left[\sum_{i=1}^N \log_2\left(1+\frac{\mathcal{P}}{N_t}\lambda_i\right)\right]\\
&=&\!\!\! N\mathbb{E}\left[\log_2\left(1+\frac{\mathcal{P}}{N_t}\lambda\right)\right].
\end{eqnarray}
Here $\mathcal{P}$ constrains the total transmitted power as $\mathbb{E}[\mathbf{x}^\dag \mathbf{x}]\le \mathcal{P}$. Since $P(\{\lambda\};\Omega;\tau)$ is symmetric in all the eigenvalues the ergodic capacity can be obtained from the level density $R_1(\lambda;\Omega;\tau)$ as
\begin{equation}
\label{chcap}
\mathcal{C} =\int_0^\infty \log_2\left(1+\frac{\mathcal{P}}{N_t}\lambda\right) R_1(\lambda;\Omega;\tau)\, d\lambda.
\end{equation}
It is possible to carry out this integral explicitly by considering the expansion of associated Laguerre polynomials in (\ref{dens}), as done for the Rayleigh fading in \cite{SL1}. The final result would involve a triple summation. We present here the results by carrying out the integration in (\ref{chcap}) numerically.

\begin{figure}[!t]
\centering
\includegraphics[width=3in]{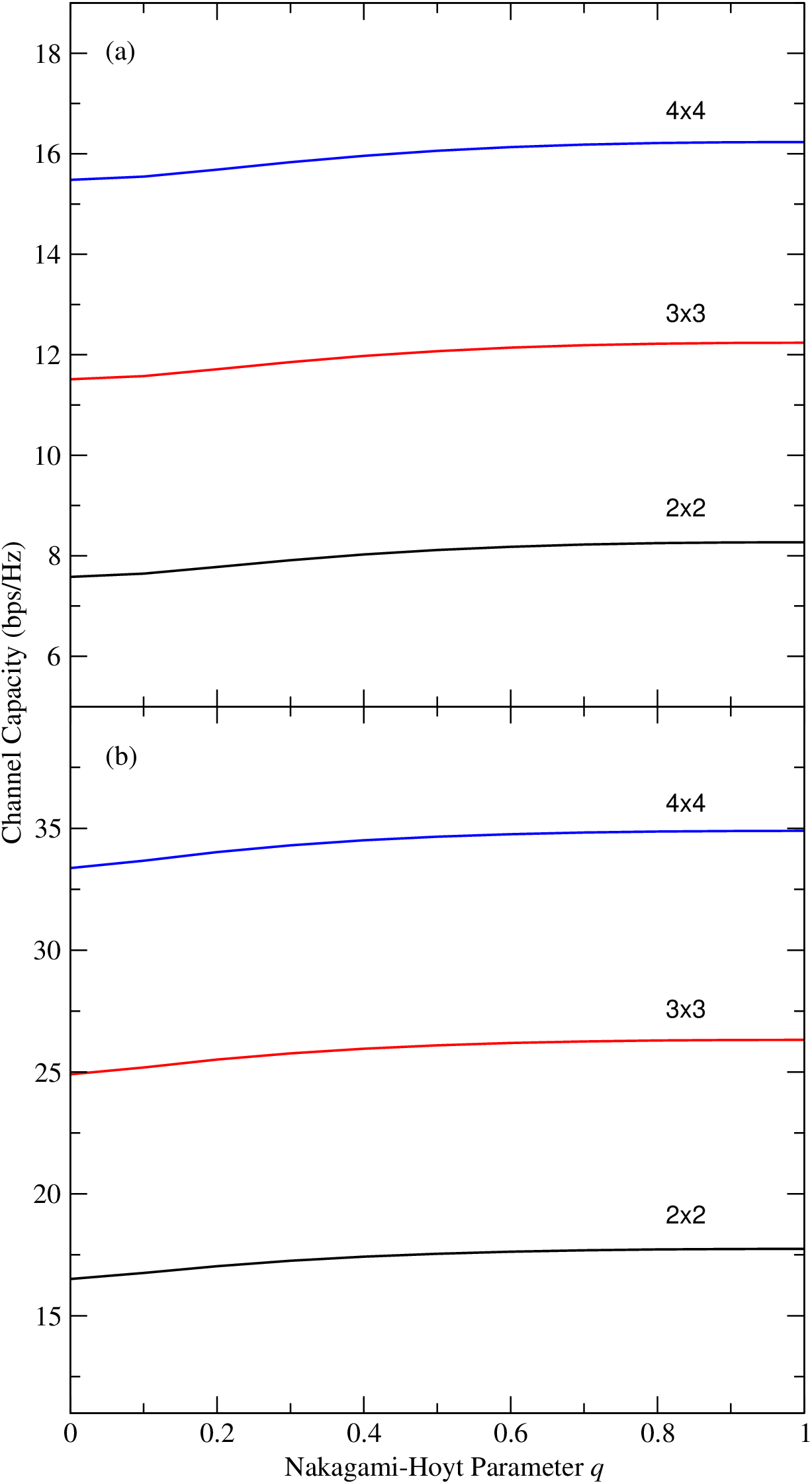}
\caption{Channel Capacity vs $q$ at (a) $\mathcal{P}=15\mbox{dB}$ and (b) $\mathcal{P}=30\mbox{dB}$ for several $N_t=N_r$ values with $\Omega=1$.}
\label{ChCapq}
\end{figure}

\begin{figure}[!t]
\centering
\includegraphics[width=3in]{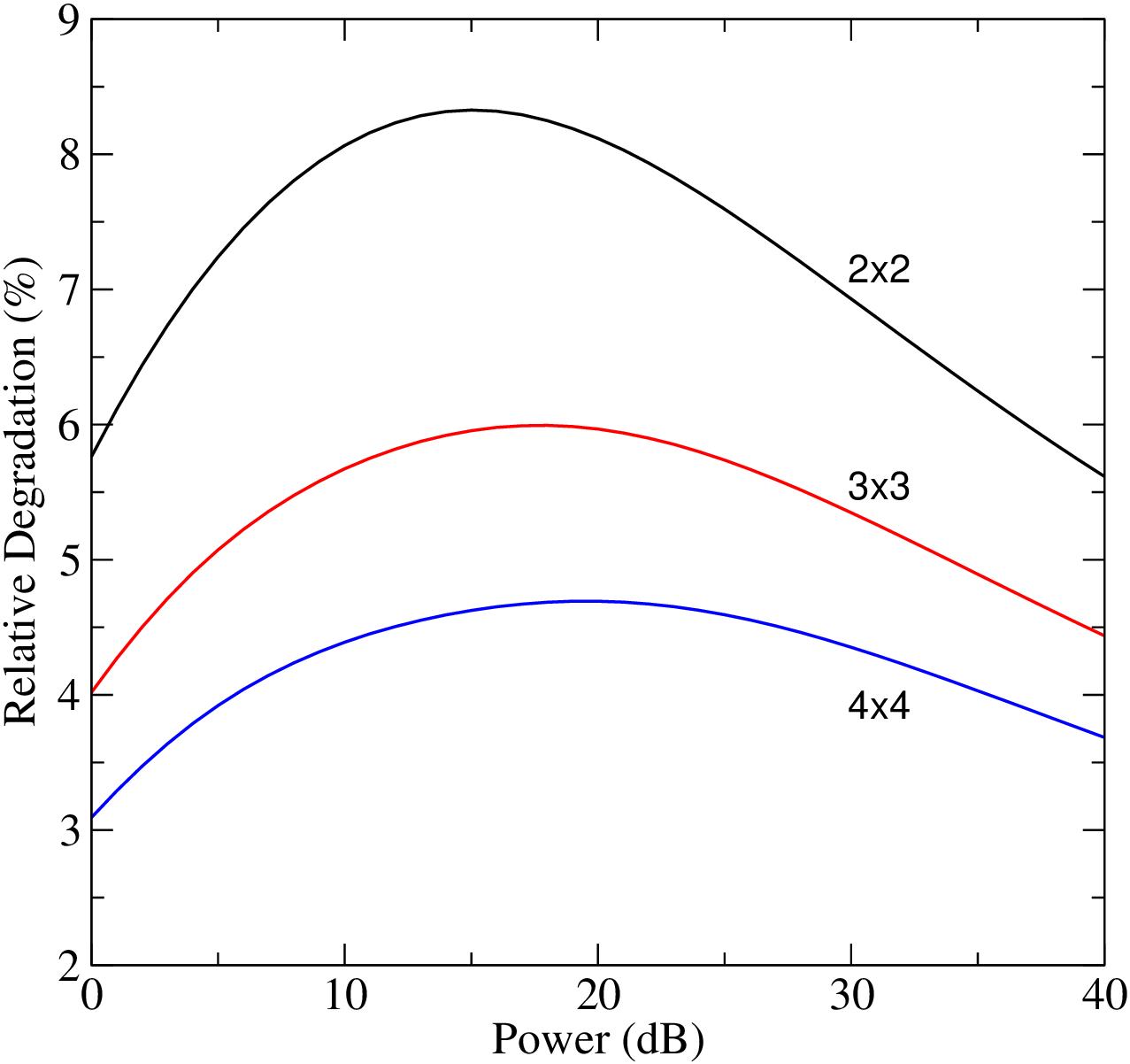}
\caption{Relative degradation vs $\mathcal{P}$ for several values of $N_t=N_r$ with $\Omega=1$.}
\label{Degrad}
\end{figure}

Fig. \ref{ChCap} shows the channel capacity for various values of $N_t=N_r$ in the extreme cases, i.e. for $q=0$ and $q=1$. The result using asymptotic density is also given. The effect of the Nakagami-Hoyt parameter $q$ on the channel capacity is shown in Fig. \ref{ChCapq} for several $N_t,N_r$ values at $\mathcal{P}=$ 15dB and 30dB. The degradation in channel capacity can be clearly seen as $q$ decreases from 1 to 0. For example, at $\mathcal{P}=15\mbox{dB}$ the channel capacity is found to degrade upto 8.33\%, 5.96\% and 4.63\% in $N_t=N_r=2,3,4$ cases respectively. Fig. \ref{Degrad} shows the relative degradation in the channel capacity, $1-\mathcal{C}(q=0)/\mathcal{C}(q=1)$, as a function of $\mathcal{P}$.

\section{Conclusion}
\label{secConc}

We have shown that the problem of Nakagami-$q$ fading in MIMO channel can be effectively and exactly handled by considering the Laguerre crossover ensemble of random matrices. Employing the techniques introduced in \cite{SKP1} we have given exact expressions for the JPD as well as density correlation functions of all orders. The level density of the eigenvalues is presented in terms of a series consisting of associated Laguerre polynomials and is used in the calculation of Shannon channel capacity. The higher order correlation functions will be useful in studying the distribution of the channel capacity \cite{SS,GWB,KHY,YMLJ}.

The present formalism can also be used to solve the problem of uncorrelated Rician fading in MIMO channels, as indicated in Appendix A. It would be of interest to extend it to deal with the cases of correlated channels and also to consider the phenomena of keyhole or pinhole channels which arise because of degenerate channels \cite{MS,FG2,SL2}.

\appendices

\section{Brownian Motion of Laguerre Ensembles}

For the Brownian motion model of the Laguerre ensembles it is convenient to work in terms of the parameter $\tau$ which serves as a fictitious time \cite{Pandey}. It is related to the parameter $q$ as given in (\ref{tau}). At $\tau=0$ the matrix $\mathbf{H}$ will have only real Gaussian elements, whereas for $\tau\rightarrow\infty$ it will have complex Gaussian elements with equal variances for the real and imaginary parts. In the following, without loss of any generality, we take $\mathbf{H}$ to be $M\times N$ dimensional with $M\ge N$.

The matrix model given by (\ref{Hmat}) is \textit{statistically equivalent} to the following matrix model
\begin{equation}
\label{Ht}
\widetilde{\mathbf{H}}(\tau)=e^{-\tau/2}\Omega^{1/2}\mathbf{A} + \left(\frac{1-e^{-\tau}}{2}\right)^{1/2}\Omega^{1/2}\mathbf{V},
\end{equation}
where $\mathbf{A}$ belongs to the ensemble of real matrices with iid Gaussian elements having zero mean and unit variance. $\mathbf{V}$ is a member of the ensemble of complex matrices having iid Gaussian entries with zero mean and unit variances for both real and imaginary parts. 
At $\tau+\delta\tau$ we have
\begin{equation}
\label{Htdt}
\widetilde{\mathbf{H}}(\tau+\delta\tau)=e^{-\tau/2-\delta\tau/2}\Omega^{1/2}\mathbf{A} + \left(\frac{1-e^{-\tau-\delta\tau}}{2}\right)^{1/2}\Omega^{1/2}\mathbf{V}.
\end{equation}
As far as eigenvalue properties are concerned, ensemble of matrices described by (\ref{Htdt}) is again \textit{statistically equivalent} to the ensemble of matrices defined by
\begin{equation}
\label{Hbrown}
\widehat{\mathbf{H}}(\tau+\delta\tau)=a_0 (\widetilde{\mathbf{H}}(\tau) + \epsilon\widetilde{\mathbf{V}}),
\end{equation}
where $\widetilde{\mathbf{V}}$ belongs to the same ensemble as $\mathbf{V}$, and $a_0$ and $\epsilon$ are given by
\begin{equation}
\label{a0}
a_0=e^{-\delta\tau/2}= 1-\frac{1}{2}\delta\tau, 
\end{equation}
and
\begin{equation}
\label{eps}
\epsilon=\left(\frac{e^{\delta\tau}-1}{2}\right)^{1/2}\Omega^{1/2}=\left(\frac{\Omega}{2}\right)^{1/2}(\delta\tau)^{1/2}. 
\end{equation}
Here in (\ref{a0}) and (\ref{eps}) as well as other equations below the expressions are correct upto order $\delta\tau$. Considering the singular value decomposition $\mathbf{U}_1^\dag \widetilde{\mathbf{H}}(\tau) \mathbf{U}_2=\mathbf{X}(\tau)$, we obtain from (\ref{Hbrown}) 
\begin{equation}
\mathbf{X}(\tau+\delta\tau)=\left(1-\frac{1}{2}\delta\tau\right)\mathbf{X}(\tau)+\left(\frac{\Omega}{2}\right)^{1/2}(\delta\tau)^{1/2}\widehat{\mathbf{V}}.
\end{equation}
Here $\mathbf{X}(\tau)$ is $M\times N$ dimensional real rectangular diagonal matrix. $\widehat{\mathbf{V}}(=\mathbf{U}_1^\dag \widetilde{\mathbf{V}}\mathbf{U}_2)$ has the same statistical properties as that of $\widetilde{\mathbf{V}}$. Thus we have 
\begin{eqnarray}
\nonumber
&& \mathbf{X}^T(\tau+\delta\tau)\mathbf{X}(\tau+\delta\tau)\equiv(1-\delta\tau) \mathbf{X}^T(\tau)\mathbf{X}(\tau)~~~~~~~~\\
&+&\!\!\!\!\big(\mathbf{X}^T(\tau)\widehat{\mathbf{V}}+\widehat{\mathbf{V}}^\dag \mathbf{X}(\tau)\big)\Big(\frac{\Omega}{2}\delta\tau\Big)^{1/2}+\frac{\Omega}{2}\widehat{\mathbf{V}}^\dag \widehat{\mathbf{V}} \delta\tau.
\end{eqnarray}
Note that the singular values of $\widetilde{\mathbf{H}}$ are square roots of eigenvalues $\{\lambda(\tau)\}$ of $\widetilde{\mathbf{H}}^\dag \widetilde{\mathbf{H}}$. As mentioned above the eigenvalues $\{\lambda\}$ are statistically equivalent to those of $\mathbf{W}=\mathbf{H}^\dag\mathbf{H}$. Similarly the eigenvalues $\{\lambda(\tau+\delta\lambda)\}$ correspond to the squares of the singular values of $\widehat{\mathbf{H}}$.
Applying the second-order perturbation theory we get,
\begin{eqnarray}
\nonumber
&&\delta \lambda_j(\tau)\equiv\lambda_j(\tau+\delta\tau)-\lambda_j(\tau)~~~~~~~~~~~~~~~~~~~~~~~~~~~~~~~~~~~~~\\
\nonumber
&&\!\!\!\!\!=-\lambda_j(\tau)\delta\tau+(\lambda_j(\tau))^{1/2}\Big(\widehat{V}_{jj}(\tau)+\widehat{V}_{jj}^{*}(\tau)\Big)\Big(\frac{\Omega}{2}\delta\tau\Big)^{1/2}\\
\nonumber
&&\!\!\!\!\!+\sum_{k=1}^{M}\widehat{V}_{kj}^{*}(\tau)\widehat{V}_{kj}(\tau)\,\Big(\frac{\Omega}{2}\delta\tau\Big)\\
\nonumber
&&\!\!\!\!\!+\!\!\!\!\sum_{k(\neq j)=1}^N\!\!\!\frac{|(\lambda_j(\tau))^{1/2}\widehat{V}_{jk}(\tau)+(\lambda_k(\tau))^{1/2}\widehat{V}_{kj}^{*}(\tau)|^2}{\lambda_j(\tau)-\lambda_k(\tau)}\,\Big(\frac{\Omega}{2}\delta\tau\Big).
\end{eqnarray}
\begin{equation}
~
\end{equation}
Since
\begin{equation}
\mathbb{E}[\widehat{V}_{jk}]=0,~~~~ \mathbb{E}[\widehat{V}_{jk}^{*}\widehat{V}_{lm}]=2\delta_{jl}\delta_{km},
\end{equation}
we get, for the eigenvalues $\{\lambda\}$, the following moments
\begin{equation}
\label{m1}
\mathbb{E}[\delta \lambda_j]=\bigg(-\lambda_j+(M-N+1)\Omega\,+\!\!\sum_{k(\neq j)=1}^N\frac{2\Omega \lambda_j}{\lambda_j-\lambda_k}\bigg)\delta \tau,
\end{equation}
and
\begin{equation}
\label{m2}
\mathbb{E}[\delta \lambda_j \delta \lambda_k]=2\Omega\,\lambda_j\,\delta_{jk}\,\delta\tau.
\end{equation}
Using these we can construct the Fokker-Planck equation for the JPD of eigenvalues \cite{RM,Dys,Pandey,Rskn},
\begin{equation}
\label{FP}
\frac{\partial P}{\partial \tau}=-\mathcal{L}P,
\end{equation}
where $\mathcal{L}$ is the Fokker-Planck operator given by
\begin{eqnarray}
\nonumber
\mathcal{L}=-\Omega \sum_j\Big[\frac{\partial^2}{\partial \lambda_j^2}\lambda_j-\frac{\partial}{\partial \lambda_j}\Big\{(M-N+1)-\frac{\lambda_j}{\Omega}\\
+\sum_{k(\neq j)}\frac{2\lambda_j}{\lambda_j-\lambda_k}\Big\}\Big].
\end{eqnarray}
The equilibrium density corresponds to LUE (Rayleigh) and can be verified by setting $\mathcal{L}P_{\text{eq}}=0$.

Using the similarity transformation\footnote{The same transformation applied to a more general ensemble with real and quaternion-real matrices leads to a Calogero-Sutherland type of Hamiltonian \cite{Pandey}.} $\xi=P_{\text{eq}}^{-1/2}P$, we map (\ref{FP}) to a Schr\"odinger equation in imaginary time $i\tau$, We get
\begin{equation}
\label{schro}
\frac{\partial \xi}{\partial \tau}=-\mathcal{H} \xi.
\end{equation}
Here $\mathcal{H}=P_{\text{eq}}^{-1/2}\mathcal{L}P_{\text{eq}}^{1/2}$ describes a system of $N$ non-interacting fermions and is given by
\begin{equation}
\label{CSH}
\mathcal{H}=\sum_{j=1}^N\mathcal{H}_{\lambda_j}-\frac{1}{2}N(N-1),
\end{equation}
where $\mathcal{H}_\lambda$ is the single particle Hamiltonian,
\begin{eqnarray}
\nonumber \mathcal{H}_\lambda=-\Omega \bigg[\lambda\frac{\partial^2}{\partial \lambda^2}+\frac{\partial}{\partial \lambda}- \frac{\lambda}{4\Omega^2}-\frac{1}{4}\frac{(M-N)^2}{\lambda}\\
+\frac{(M-N+1)}{2\Omega}\bigg].
\end{eqnarray}
The eigenfunctions of $\mathcal{H}_{\lambda}$ are the weighted associated Laguerre polynomials $[(\lambda/\Omega)^{M-N}e^{-\lambda/\Omega}]^{1/2}L_n^{(M-N)}(\lambda/\Omega)$ with corresponding eigenvalue $n$, where $n=0,1,2,...~$. The eigenfunction $\xi$ of $\mathcal{H}$ is thus the Slater determinant comprising these weighted polynomials. Also note that the ground state energy of $\mathcal{H}$ is zero.

The JPD of eigenvalues at arbitrary $\tau$ can be obtained as
\begin{equation}
\label{PXt-1}
P(\{\lambda\},\tau)=e^{-\mathcal{L}\tau}P(\{\lambda\},0)=P_{\text{eq}}^{1/2}e^{-\mathcal{H}\tau}P_{\text{eq}}^{-1/2}P(\{\lambda\},0).
\end{equation} 
This expression is equivalent to
\begin{equation}
\label{PXt-2}
P(\{\lambda\},\tau)=\int_{0}^\infty\!\!\! d\gamma_1 \cdots\!\!\!\int_{0}^\infty \!\!\!d\gamma_N \boldsymbol{P}(\{\lambda\};\tau|\{\gamma\};0)P(\{\gamma\},0),
\end{equation}
where $\boldsymbol{P}(\{\lambda\};\tau|\{\gamma\};0)$ is the conditional JPD (or Green's function or propagator) which gives the probability of finding the eigenvalues $\{\lambda\}$ at $\tau$ when the initial eigenvalues are $\{\gamma\}$. We have \cite{Pandey}
\begin{eqnarray}
\nonumber
\boldsymbol{P}(\{\lambda\};\tau|\{\gamma\};0)=\left(\frac{P_{\text{eq}}(\{\lambda\})}{P_{\text{eq}}(\{\gamma\})}\right)^{1/2}~~~~~~~~~~~~~~~~~~~\\
~~~~\times\sum_{n}e^{-\mathcal{E}_{n}\tau}\xi_{n}(\{\lambda\})\xi_{n}(\{\gamma\}),
\end{eqnarray}
where the sum is over all the states. $\mathcal{E}_n$ is the eigenvalue corresponding to the state $\xi_{n}(\{\lambda\})$. 
Substituting the Slater determinant form of $\xi$ we obtain
\begin{eqnarray}
\label{green}
\nonumber
\boldsymbol{P}(\{\lambda\};\tau|\{\gamma\};0)=\frac{1}{\Omega^N N!}\frac{\Delta_N(\{\lambda\})}{\Delta_N(\{\gamma\})}\prod_{i=1}^N w_{2a+1}\Big(\frac{\lambda_i}{\Omega}\Big)\\
\times e^{N(N-1)\tau/2}\mathrm{det}\left[f\Big(\frac{\lambda_j}{2\Omega},\frac{\gamma_k}{2\Omega};\tau\Big)\right]_{j,k=1,...,N},
\end{eqnarray}
where \cite{Szego}
\begin{eqnarray}
\nonumber
\label{fxyt}
f(x,y;\tau)=\sum_{\mu=0}^\infty\frac{e^{-\mu\tau}}{\alpha_\mu^2}L_\mu^{(2a+1)}(2x)L_\mu^{(2a+1)}(2y)~~~~~~\\
=\frac{e^{-2(x+y)e^{-\tau}/(1-e^{-\tau})}}{(1-e^{-\tau})(-4xye^{-\tau})^{a+1/2}}\,J_{2a+1}\left(\frac{4(-xye^{-\tau})^{1/2}}{1-e^{-\tau}}\right),
\end{eqnarray}
$J_k(z)$ is the Bessel function and
\begin{equation}
\alpha_\mu=\bigg[\frac{\Gamma(\mu+2a+2)}{\Gamma(\mu+1)}\bigg]^{1/2}.
\end{equation}
Eq. (\ref{green}) is similar to the expression obtained after integral over unitary group \cite{SMM}. In the present problem we use it to solve the Nakagami-$q$ fading problem. However it can be used to tackle other fading distributions also. For example, the uncorrelated Rician case can be solved by taking $\mathbf{A}$ in (\ref{Ht}) as a fixed matrix instead of belonging to an ensemble of real Gaussian matrices and properly adjusting the values of $\Omega$ and $\tau$. 

\section{JPD for LOE-LUE Crossover}

To find the JPD for LOE-LUE crossover we need to substitute for $P(\{\lambda\};0)$ the JPD for the LOE. We can use either (\ref{PXt-1}) or (\ref{PXt-2}) to accomplish this task. It turns out that it is advantageous to use (\ref{PXt-1}) as it leads to the final result in fewer steps \cite{SKP1}. This is done by expressing (\ref{JPDzero}) in terms of a Pfaffian. 

The Pfaffian of a $2m \times 2m$ antisymmetric matrix $\mathbf{B}$ is defined as \cite{RM}
\begin{equation}
\label{Pf}
\mbox{Pf}[\mathbf{B}]=\sum_{\mathrm{p}} \sigma_{\mathrm{p}} B_{i_1,i_2}B_{i_3,i_4}\cdots B_{i_{2m-1},i_{2m}}.
\end{equation}
The sum in (\ref{Pf}) is over all permutations $$\mbox{p}=\begin{pmatrix}1 & 2 & ... & 2m\\i_1&i_2&...&i_{2m}\end{pmatrix}$$ with the restrictions $i_1<i_2,~i_3<i_4,~...,~i_{2m-1}<i_{2m}$; $i_1<i_3<...<i_{2m-1}$ and $\sigma_{\mathrm{p}}$ is sign of the permutation. Also, Pfaffian is related to the determinant as 
\begin{equation}
\det[\mathbf{B}]=(\mbox{Pf}[\mathbf{B}])^2.
\end{equation}

Thus we can write $P(\{\lambda\};0)$ as
\begin{equation}
 P(\{\lambda\};0)\propto\Delta_N \mbox{Pf}[F_{j,k}^{(0)}]\prod_{l=1}^N w_a\left(\frac{\lambda_l}{2\Omega}\right),
\end{equation}
where $F_{j,k}^{(0)}$ is given using (\ref{Fjk_e}), (\ref{Fjk_o}), (\ref{Gxy0}) and (\ref{OMGx0}).

Using (\ref{CSH}) and (\ref{PXt-1}) we get
\begin{eqnarray}
\nonumber
P(\{\lambda\},\tau)\propto e^{N(N-1)\tau/2}\Delta_N \left(\bigg(\prod_{i=1}^N \boldsymbol{O}_{\lambda_i}\bigg) \mbox{Pf}[F_{j,k}^{(0)}]\right)\\
\times\prod_{l=1}^N w_a\left(\frac{\lambda_l}{2\Omega}\right),~~~~~~~~~~ 
\end{eqnarray}
where we have introduced the one-body operator $\boldsymbol{O}_\lambda$ given by
\begin{equation}
 \boldsymbol{O}_\lambda=\frac{\sqrt{w_{2a+1}(\frac{\lambda}{\Omega})}}{w_a(\frac{\lambda}{2\Omega})}e^{-\mathcal{H}_{\lambda}\tau}\frac{w_a(\frac{\lambda}{2\Omega})}{\sqrt{w_{2a+1}(\frac{\lambda}{\Omega})}}.
\end{equation}
We will also need 
\begin{equation}
(\boldsymbol{O}_\lambda^\dag)^{-1}=\frac{w_a(\frac{\lambda}{2\Omega})}{\sqrt{w_{2a+1}(\frac{\lambda}{\Omega})}} e^{\mathcal{H}_{\lambda}\tau} \frac{\sqrt{w_{2a+1}(\frac{\lambda}{\Omega})}}{w_a(\frac{\lambda}{2\Omega})}.
\end{equation}
The operator $\boldsymbol{O}_\lambda$ has eigenfunctions $w_{a+1}(\frac{\lambda}{2\Omega})L_\mu^{(2a+1)}(\frac{\lambda}{\Omega})$ with eigenvalues $e^{-\mu\tau}$, whereas $(\boldsymbol{O}_\lambda^\dag)^{-1}$ has eigenfunctions $w_a(\frac{\lambda}{2\Omega})L_\mu^{(2a+1)}(\frac{\lambda}{\Omega})$ with eigenvalues $e^{\mu\tau}$. Eq. (\ref{JPD}) is then obtained using the expansion of Pfaffian with
\begin{equation}
\label{Gevol}
\mathcal{G}^{(\tau)}(x,y)=\boldsymbol{O}_x \boldsymbol{O}_y \mathcal{G}^{(0)}(x,y),
\end{equation}
and
\begin{equation}
\label{OMGevol}
\omega^{(\tau)}(x)=\boldsymbol{O}_x\omega^{(0)}(x).
\end{equation}

The antisymmetric function $\mathcal{G}^{(0)}(x,y)=\frac{1}{2}\mbox{sgn}(x-y)$ can be expanded in terms of the functions dual to the weighted skew-orthogonal polynomials \cite{PG,GP,SKP1}, as defined in Appendix C. We have
\begin{equation}
\label{Gexp}
 \mathcal{G}^{(0)}(x,y)=\sum_{\mu=0}^\infty[\psi_{2\mu}^{(0)}(x)\psi_{2\mu+1}^{(0)}(y)-\psi_{2\mu+1}^{(0)}(x)\psi_{2\mu}^{(0)}(y)].
\end{equation}
We use $\tau=0$ expressions for skew-orthogonal polynomials and their dual functions from \cite{PG,GP}. The $\tau$-dependent counterparts given in Appendix C are their generalizations for the crossover problem \cite{SKP1}.

The expansion of $\omega^{(0)}(x)=\frac{1}{2}$ is obtained by writing it as $\frac{1}{2}\int_0^\infty \delta(x-y)\,dy$ and using
\begin{equation}
\delta(x-y)=w_{a+1}(x)w_a(y)\sum_{\mu=0}^\infty \frac{1}{\alpha_\mu^2} L_\mu^{(2a+1)}(2x)L_\mu^{(2a+1)}(2y).
\end{equation}
The integration over this expansion of delta-function is accomplished by recursively using the following important relation which holds for nonnegative integers $\mu$ \cite{GP},
\begin{eqnarray}
\label{imp}
\nonumber
 \frac{d}{dx}[w_{a+1}(x)L_\mu^{(2a+1)}(2x)]\!\!\!&=&\!\!\!\frac{1}{2}w_a(x)[A_{\mu+1}L_{\mu+1}^{(2a+1)}(2x)\\
&& -B_{\mu-1}L_{\mu-1}^{(2a+1)}(2x)],
\end{eqnarray}
with $A_\mu=\mu, B_\mu=\mu+2a+2, B_{-1}=0$. We find
\begin{equation}
\label{san}
\int_0^\infty w_a(y)L_\mu^{(2a+1)}(2y)\,dy
=\begin{cases}
 \dfrac{\Gamma(\frac{\mu}{2}+a+1)}{\Gamma(\frac{\mu}{2}+1)},& \mu\mbox{ even},\\
 ~~~~~0, & \mu\mbox{ odd}.
 \end{cases} 
\end{equation}
Thus we get infinite series representation for $\omega^{(0)}(x)$. Application of operator $\boldsymbol{O}$ on $\tau=0$ results, as given by (\ref{Gevol}) and (\ref{OMGevol}), then leads to the expressions (\ref{Gxyt}), (\ref{OMGxt}) of $\mathcal{G}$ and $\omega$ valid for arbitrary $\tau$.

\section{Skew-orthogonal polynomials and density correlation functions}

We give here the results for $\Omega=1/2$ and, by a simple rescaling, obtain results for arbitrary $\Omega$.

The weighted skew-orthogonal polynomials $\phi_{j}^{(\tau)}(x)$ and their dual functions $\psi_{j}^{(\tau)}(x)$ satisfy the following relations \cite{RM,PG,GP,SKP1,SKP2}:
\begin{equation}
\int_{0}^\infty \phi_j^{(\tau)}(x)\psi_k^{(\tau)}(x)\,dx=\mathcal{Z}_{j,k},
\end{equation}
\begin{equation}
\psi_j^{(\tau)}(x)=\int_0^\infty \mathcal{G}^{(\tau)}(x,y) \phi_j^{(\tau)}(y)\,dy. 
\end{equation}
Here
\begin{equation}
\mathcal{Z}_{j,k}=\begin{cases}
   ~~1, & j \mbox{ even and } k=j+1,\\
	 -1, & j \mbox{ odd and } k=j-1,\\
	 ~~0, & \mbox{ otherwise}.
        \end{cases}
\end{equation}
When $N$ is odd we have an extra condition:
\begin{equation}
\int_0^\infty \omega^{(\tau)}(x)\phi_{j}^{(\tau)}(x)\,dx=\delta_{j,N-1}.
\end{equation}
The $\tau$-dependent expressions for $\phi$ and $\psi$ are obtained by operating $(\boldsymbol{O}_x^\dag)^{-1}$ and $\boldsymbol{O}_x$ on their $\tau=0$ counterparts as \cite{SKP1}
\begin{equation}
\phi_\mu^{(\tau)}(x)=(\boldsymbol{O}_x^\dag)^{-1}\phi_\mu^{(0)}(x), 
\end{equation}
and
\begin{equation}
\psi_\mu^{(\tau)}(x)=\boldsymbol{O}_x\,\psi_\mu^{(0)}(x).
\end{equation}

For even $N$ we find
\begin{eqnarray}
\label{Ephie}
 \phi_{2\mu}^{(\tau)}(x)\!\!\!&=&\!\!\!\frac{2^{a+1/2}}{\alpha_{2\mu}}w_a(x)e^{2\mu\tau}L_{2\mu}^{(2a+1)}(2x),~~~~~~~~~~~~~~~~~
\end{eqnarray}
\begin{eqnarray}
\label{Epsie} 
\nonumber
\psi_{2\mu}^{(\tau)}(x)\!\!\!&=&\!\!\!-\frac{2^{a+3/2}}{\alpha_{2\mu}}w_{a+1}(x)\sum_{\nu=\mu}^\infty\big[e^{-(2\nu+1)\tau}~~~~~~~~~~~~~~~\\
&&\times\frac{\Gamma(\nu+1)\Gamma(\mu+a+1)}{2\,\Gamma(\nu+a+2)\Gamma(\mu+1)}L_{2\nu+1}^{(2a+1)}(2x)\big],
\end{eqnarray}
\begin{eqnarray}
\label{Ephio} 
\nonumber      
 \phi_{2\mu+1}^{(\tau)}(x)\!\!\!&=&\!\!\!\frac{2^{a+1/2}}{\alpha_{2\mu}}w_a(x)\big[e^{(2\mu+1)\tau}(2\mu+1)L_{2\mu+1}^{(2a+1)}(2x)\\
 &-&\!\!\!e^{(2\mu-1)\tau}(2\mu+2a+1)L_{2\mu-1}^{(2a+1)}(2x)\big],
\end{eqnarray}
\begin{eqnarray}
\label{Epsio}
 \psi_{2\mu+1}^{(\tau)}(x)=\frac{2^{a+3/2}}{\alpha_{2\mu}}w_{a+1}(x)e^{-2\mu\tau}L_{2\mu}^{(2a+1)}(2x),~~~~~~~
\end{eqnarray}
for $\mu=0,1,2,...$ .
For odd $N$ these are given by 
\begin{eqnarray}
\label{Ophie} 
 \phi_{2\mu}^{(\tau)}(x)\!\!\!&=&\!\!\!\frac{2^{a+1/2}}{\alpha_{2\mu+1}}w_a(x)e^{(2\mu+1)\tau}L_{2\mu+1}^{(2a+1)}(2x),~~~~~~~~~~~~
\end{eqnarray}
\begin{eqnarray}
\label{Opsie} 
\nonumber
\psi_{2\mu}^{(\tau)}(x)\!\!\!&=&\!\!\!\frac{2^{a+3/2}}{\alpha_{2\mu+1}}w_{a+1}(x)\sum_{\nu=0}^\mu\big[e^{-2\nu\tau}~~~~~~~~~~~~~~~~~~~~\\
&&\!\!\!\times\frac{\Gamma(\nu+\frac{1}{2})\Gamma(\mu+a+\frac{3}{2})}{2\,\Gamma(\nu+a+\frac{3}{2})\Gamma(\mu+\frac{3}{2})}L_{2\nu}^{(2a+1)}(2x)\big],
\end{eqnarray}
\begin{eqnarray}
\label{Ophio} 
\nonumber
 \phi_{2\mu+1}^{(\tau)}(x)\!\!\!&=&\!\!\!\frac{2^{a+1/2}}{\alpha_{2\mu+1}}w_a(x)\big[e^{(2\mu+2)\tau}(2\mu+2)L_{2\mu+2}^{(2a+1)}(2x)\\
 &-&\!\!\!e^{2\mu\tau}(2\mu+2a+2)L_{2\mu}^{(2a+1)}(2x)\big],
\end{eqnarray}
\begin{eqnarray}
\label{Opsio} 
\psi_{2\mu+1}^{(\tau)}(x)=\frac{2^{a+3/2}}{\alpha_{2\mu+1}}w_{a+1}(x)e^{-(2\mu+1)\tau}L_{2\mu+1}^{(2a+1)}(2x),
\end{eqnarray}
for $\mu=0,1,2,...,(N-3)/2$. The unpaired $\phi$ for odd $N$ is given by
\begin{equation}
\phi_{N-1}^{(\tau)}(x)=2w_a(x)e^{(N-1)\tau}\frac{\Gamma(\frac{N+1}{2})}{\Gamma(\frac{N+2a+1}{2})}L_{N-1}^{(2a+1)}(2x),
\end{equation}
with its dual
\begin{eqnarray}
\nonumber
\psi_{N-1}^{(\tau)}(x)\!\!\!&=&\!\!\!-2w_{a+1}(x)\!\!\!\sum_{\nu=\frac{(N-1)}{2}}^\infty\big[e^{-(2\nu+1)\tau}\\
&&\times\frac{\Gamma(\nu+1)}{\Gamma(\nu+a+2)}L_{2\nu+1}^{(2a+1)}(2x)\big].
\end{eqnarray}

Using these weighted skew-orthogonal polynomials and their dual functions we define the following two-point kernels
\begin{eqnarray}
\label{Sxyt}
\nonumber
S_N^{(\tau)}(x,y)\!\!\!&=&\!\!\!\sum_{\mu=0}^{\frac{(N-c)}{2}-1}\!\!\!\left[\phi_{2\mu}^{(\tau)}(x)\psi_{2\mu+1}^{(\tau)}(y)-\phi_{2\mu+1}^{(\tau)}(x)\psi_{2\mu}^{(\tau)}(y)\right]\\
\nonumber
&+&\!\!\!c\,\phi_{N-1}^{(\tau)}(x)\omega^{(\tau)}(y)\\
&=&\!\!\!(\boldsymbol{O}_x^\dag)^{-1} \boldsymbol{O}_y \, S_N^{(0)}(x,y),
\end{eqnarray}
\begin{eqnarray}
\label{Axyt}
\nonumber
A_N^{(\tau)}(x,y)\!\!\!&=&\!\!\!\sum_{\mu=0}^{\frac{(N-c)}{2}-1}\!\!\!\left[\phi_{2\mu+1}^{(\tau)}(x)\phi_{2\mu}^{(\tau)}(y)-\phi_{2\mu}^{(\tau)}(x)\phi_{2\mu+1}^{(\tau)}(y)\right]\\
&=&\!\!\!(\boldsymbol{O}_x^\dag)^{-1} (\boldsymbol{O}_y^\dag)^{-1} A_N^{(0)}(x,y),
\end{eqnarray}
\begin{eqnarray}
\label{Bxyt}
\nonumber
B_N^{(\tau)}(x,y)\!\!\!&=&\!\!\!\sum_{\mu=\frac{(N-c)}{2}}^{\infty}\!\!\!\left[\psi_{2\mu+1}^{(\tau)}(x)\psi_{2\mu}^{(\tau)}(y)-\psi_{2\mu}^{(\tau)}(x)\psi_{2\mu+1}^{(\tau)}(y)\right]\\
\nonumber
&+&c\,[\psi_{N-1}^{(\tau)}(x)\omega^{(\tau)}(y)-\psi_{N-1}^{(\tau)}(y)\omega^{(\tau)}(x)]\\
&=&\!\!\!\boldsymbol{O}_x\boldsymbol{O}_y B_N^{(0)}(x,y),
\end{eqnarray}
Recall that $c=N\mbox{ (mod 2)}$. Using Dyson's theorems \cite{RM} we can write the $n$-level correlation function defined by (\ref{Rn}) as a quaternion determinant\footnote{Just like the Pfaffian, the square of the quaternion determinant of a ``self-dual'' matrix is the ordinary determinant of the corresponding $2N\times 2N$ matrix. See \cite{RM} for details.} involving these kernels, viz.
\begin{equation}
\label{RnQdet}
\mathcal{R}_n(x_1,...,x_n;\tau)=\mbox{Qdet}[\sigma^{(\tau)}(x_j,x_k)]_{j,k=1,..,n},
\end{equation}
where $\sigma^{(\tau)}(x,y)$ is given by
\begin{equation}
\sigma^{(\tau)}(x,y)
=\begin{bmatrix}
    S_N^{(\tau)}(x,y) & A_N^{(\tau)}(x,y)\\
    B_N^{(\tau)}(x,y) & S_N^{(\tau)}(y,x)
\end{bmatrix}.
\end{equation}
In particular the level density is obtained as 
\begin{equation}
\label{R1}
\mathcal{R}_1(x;\tau)=S_N^{(\tau)}(x,x). 
\end{equation}
The correlation function $R_n(\lambda_1,...,\lambda_n;\Omega;\tau)$ of (\ref{dens}) is obtained by rescaling as
\begin{equation}
\label{scale}
 R_n(\lambda_1,...,\lambda_n;\Omega;\tau)=\frac{1}{(2\Omega)^n}\mathcal{R}_n\Big(\frac{\lambda_1}{2\Omega},...,\frac{\lambda_n}{2\Omega};\tau\Big).
\end{equation}

\section{Proofs of equations (\ref{Gxyt}), (\ref{dens}) and (\ref{Dzero})}

$\mathcal{G}^{(\tau)}(x,y)$ is obtained by direct substitution of $\psi^{(\tau)}$s in $\tau$-dependent analogue of (\ref{Gexp}) which is similar to (\ref{Sxyt})-(\ref{Bxyt}) \footnote{Or equivalently by first writing the expansion of $\mathcal{G}^{(0)}(x,y)$ and then operating $\boldsymbol{O_x}\boldsymbol{O_y}$.}. The first expression in (\ref{Gxyt}) follows by using (\ref{Epsio}) and (\ref{Epsie}), whereas the second expression is obtained by using (\ref{Opsio}) and (\ref{Opsie}). As expected these two expressions are equal and follow from each other by changing the order of summation.

Now we outline the proof of (\ref{dens}). For even $N$ we get
\begin{eqnarray}
\label{Epp1}
\nonumber
\sum_{\mu=0}^{\frac{N}{2}-1}\phi_{2\mu}^{(\tau)}(x)\psi_{2\mu+1}^{(\tau)}(y)=2^{2a+2} w_{a}(x)w_{a+1}(y)\\
\times\sum_{\mu=0}^{\frac{N}{2}-1}\frac{1}{\alpha_{2\mu}^2}L_{2\mu}^{(2a+1)}(2x)L_{2\mu}^{(2a+1)}(2y),
\end{eqnarray}
and after some algebraic manipulation,
\begin{eqnarray}
\label{Epp2}
\nonumber
\sum_{\mu=0}^{\frac{N}{2}-1}\phi_{2\mu+1}^{(\tau)}(x)\psi_{2\mu}^{(\tau)}(y)=-2^{2a+2} w_{a}(x)w_{a+1}(y)\\
\nonumber
\times\sum_{\mu=0}^{\frac{N}{2}-1}\frac{1}{\alpha_{2\mu+1}^2}L_{2\mu+1}^{(2a+1)}(2x)L_{2\mu+1}^{(2a+1)}(2y)\\
\nonumber
-2^{2a+2} w_{a}(x)w_{a+1}(y)L_{N-1}^{(2a+1)}(2x)\sum_{\nu=\frac{N}{2}}^\infty\frac{e^{-(2\nu+2-N)\tau}}{2^{2a+1}}\\
\times \frac{\Gamma(\frac{N+1}{2})\Gamma(\nu+1)}{\Gamma(\frac{N+2a+1}{2})\Gamma(\nu+a+2)}L_{2\nu+1}^{(2a+1)}(2y).
\end{eqnarray}
Thus subtracting (\ref{Epp2}) from (\ref{Epp1}) we get the kernel $S_N^{(\tau)}(x,y)$. For odd $N$ we have
\begin{eqnarray}
\nonumber
\sum_{\mu=0}^{\frac{(N-1)}{2}-1}\phi_{2\mu}^{(\tau)}(x)\psi_{2\mu+1}^{(\tau)}(y)=2^{2a+2} w_{a}(x)w_{a+1}(y)\\
\times\sum_{\mu=0}^{\frac{(N-1)}{2}-1}\frac{1}{\alpha_{2\mu+1}^2}L_{2\mu+1}^{(2a+1)}(2x)L_{2\mu+1}^{(2a+1)}(2y),
\end{eqnarray}
while
\begin{eqnarray}
\nonumber
\sum_{\mu=0}^{\frac{(N-1)}{2}-1}\phi_{2\mu}^{(\tau)}(x)\psi_{2\mu+1}^{(\tau)}(y)+ \phi_{N-1}^{(\tau)}(x)\omega^{(\tau)}(y)~~~~~~~~\\
\nonumber
=-2^{2a+2} w_{a}(x)w_{a+1}(y)\!\!\!\sum_{\mu=0}^{\frac{(N-1)}{2}}\!\!\!\frac{1}{\alpha_{2\mu}^2}L_{2\mu}^{(2a+1)}(2x)L_{2\mu}^{(2a+1)}(2y)\\
\nonumber
-2^{2a+2} w_{a}(x)w_{a+1}(y)L_{N-1}^{(2a+1)}(2x)\!\!\!\!\!\sum_{\nu=\frac{(N+1)}{2}}^\infty\!\!\frac{e^{-(2\nu+1-N)\tau}}{2^{2a+1}}\\
\nonumber
\times \frac{\Gamma(\frac{N+1}{2})\Gamma(\nu+\frac{1}{2})}{\Gamma(\frac{N+2a+1}{2})\Gamma(\nu+a+\frac{3}{2})}L_{2\nu}^{(2a+1)}(2y).\\
~ \hspace{-4cm}
\end{eqnarray}
Again $S_N^{(\tau)}(x,y)$ is obtained using (\ref{Sxyt}).

The even and odd results can be combined in a single expression as
\begin{eqnarray}
\nonumber
&&\!\!\!S_N^{(\tau)}(x,y)\\
\nonumber
&=&\!\!\!2^{2a+2}w_a(x)w_{a+1}(y)\sum_{\mu=0}^{N-1}\frac{1}{\alpha_\mu^2}L_\mu^{(2a+1)}(2x)L_\mu^{(2a+1)}(2y)\\
\nonumber
&+&\!\!\!2^{2a+2}w_a(x)w_{a+1}(y)L_{N-1}^{(2a+1)}(2x)\!\!\!\!\sum_{\nu=\frac{(N+c)}{2}}^\infty \!\!\!\frac{e^{-(2\nu+2-N-c)\tau}}{2^{2a+1}}\\
&\times&\!\!\! \frac{\,\Gamma\Big(\frac{N+1}{2}\Big)\,\Gamma(\nu+1-\frac{c}{2})}{\,\Gamma\Big(\frac{N+2a+1}{2}\Big)\,\Gamma(\nu+a+2-\frac{c}{2})}L_{2\nu+1-c}^{(2a+1)}(2y).
\end{eqnarray}
This equation generalizes the LOE result given by Widom \cite{Widom} to the Laguerre crossover ensemble. The level density in (\ref{dens}) is obtained using (\ref{R1}) and (\ref{scale}).

To prove (\ref{Dzero}) we consider the function
\begin{eqnarray}
\label{int}
\nonumber
&&\mathcal{I}_N(x)=\frac{1}{2}\int_0^\infty \mbox{sgn}(x-y)w_a(y)L_N^{(2a+1)}(2y)dy\\
\nonumber
&=&\!\!\!\frac{1}{2}\int_0^\infty \!\!\!w_a(y)L_N^{(2a+1)}(2y)dy-\int_x^\infty\!\!\! w_a(y)L_N^{(2a+1)}(2y)dy.\\
\end{eqnarray}
The first integral in second line of above equation is given using the result (\ref{san}). The second integral can be evaluated to a finite series involving incomplete gamma functions. This is accomplished by using the standard expansion of Laguerre polynomials \cite{Szego}
\begin{equation}
 L_n^{(k)}(x)=\sum_{\mu=0}^n \frac{(-1)^\mu \Gamma(n+k+1)}{\Gamma(k+\mu+1)\Gamma(n-\mu+1)} \frac{x^\mu}{\Gamma(\mu+1)},
\end{equation}
thereby giving
\begin{eqnarray}
\nonumber
 &&\!\!\!\!\!\!\mathcal{I}_N(x)=\frac{\Gamma(\frac{N}{2}+a+1)}{2\Gamma(\frac{N}{2}+1)}\delta_{c,0}~~~~~~~~~~~~~~~~~~~~~~~~~~~~~\\
 &&-\sum_{\mu=0}^N \frac{(-1)^\mu2^\mu\Gamma(N+2a+2)\Gamma(\mu+a+1,x)}{\Gamma(\mu+2a+2)\Gamma(N-\mu+1)\Gamma(\mu+1)}.
\end{eqnarray}
Now for even $N$, by using (\ref{imp}) and performing partial integration repeatedly we can obtain for $\mathcal{I}_N(x)$ an infinite series consisting of odd-order Laguerre polynomials \footnote{In this case $\mathcal{I}_N(x)$ is same as $\psi_N^{(0)}(x)$ of even $N$ case.}. This infinite series is proportional to the infinite series appearing in (\ref{dens}) for $\tau=0$. By adjusting the factors we get the representation (\ref{Dzero}) for LOE. For $N$ odd $c_1\omega^{(0)}(x)+c_2 \mathcal{I}_N(x)$, with appropriate factors $c_1$ and $c_2$, gives rise to an infinite series comprising even-order Laguerre polynomials which is proportional to the one appearing in (\ref{dens}) for $\tau=0$. Again by introducing suitable factors we obtain (\ref{Dzero}).

\section*{Acknowledgment}
The authors would like to thank G. Fraidenraich for communicating to us his unpublished work now published as \cite{FLC1}.

\end{document}